\documentclass[a4paper,11pt]{article}
\pdfoutput=1 

\usepackage{jinstpub} 
\usepackage{siunitx}   
\usepackage{wasysym} 
\usepackage{xcolor}
\usepackage[colorinlistoftodos,prependcaption]{todonotes}
\usepackage{booktabs}   
\usepackage{subcaption}   

\usepackage{xspace}
\newcommand{\mude}{{\sc Mu3e}\xspace}

\usepackage{dashrule}

\title{\boldmath Timing Properties of Blue-emitting Scintillating Fibers}

\author{A. Bravar and}
\author{Y. Demets}

\affiliation{D\'{e}partement de physique nucl\'eaire et corpusculaire,  Universit\'e de Gen\`eve,\\
24, quai Ernest-Ansermet, 1211 Gen\`eve 4, Switzerland}

\emailAdd{alessandro.bravar@unige.ch}

\date{\today}

\abstract{
Scintillating fibers are widely used in particle detectors, mainly for tracking.
We have studied the properties and have compared the performance of different blue-emitting
250~$\mu$m diameter round double-clad scintillating fibers from Kuraray and Saint-Gobain.
Here we focus on the properties of the fibers over short lengths ($< 50~{\rm cm}$),
in particular on their timing performances.
We report on the light yield and attenuation,
scintillation light decay time and the achievable time resolution.}


\begin{document}
\maketitle
\flushbottom

\section{Introduction}
\label{sec:intro}

Plastic scintillating fibers are widely used in particle detectors, mainly for tracking.
For an overview of the state-of-the-art in the use of scintillating fibers
in particle detectors before the year 2000, see for example~\cite{Ruchti}.
The scintillating fibers (SciFi) are usually assembled in ribbons consisting of several staggered layers of fibers,
resulting in low mass detectors with different adaptable geometries depending on the application.
The performance of the SciFi detector depends on the physical properties of the fibers,
for instance, the spatial resolution is directly related to the fiber's diameter
(the smaller the diameter, the higher the resolution),
while the time resolution is related to the scintillation light decay time
(the shorter the decay time, the better the time resolution, especially for low photon statistics)
and the light yield.
The fine granularity, however, also increases the number of readout channels of the optical photo-sensor.
This has led to the development of multi-anode photo-multiplier tubes (MAPMT)~\cite{Kuroda},
as an effective solution to work with a large number of channels,
and for some time, the MAPMTs have been the baseline choice for SciFi detectors.
Time resolutions of $\sim 600~{\rm ps}$ have been reported
for small size detectors ($< 5~{\rm cm} \times 5~{\rm cm}$) using 6 staggered layers of $500~\mu~{\rm m}$
diameter fibers~\cite{Agoritsas}.
The introduction of silicon photo-multipliers (SiPM), in particular of multi-channel SiPM arrays pioneered in~\cite{Beischer},
rekindled the interest in the SciFi technology opening up the fields to different applications.
When coupled to SiPM arrays, the SciFi offer reliable, robust,
and cost-effective detector assemblies, 
which can also be operated in magnetic fields.
Large SciFi tracking detectors with a spatial resolution better than $100~\mu{\rm m}$ 
have been already developed and are currently in use~\cite{LHCb}.
Scintillating fibers have been further innovated over the last years. 
One of the novelties is the introduction of a novel type of luminophores
admixed to the polystyrene matrix of the fiber core,
the so-called Nanostructured Organosilicon Luminofores (NOL)~\cite{NOL11}
with high photo-luminescence and $\sim 1~{\rm ns}$ decay time
(i.e. much shorter compared to standard POPOP wavelength shifters).
In the design of a SciFi detector, the selection of a specific fiber is particularly relevant.
When developing the SciFi timing detector for the \mude experiment~\cite{Bravar},
we realized that most of the fiber properties reported in the literature or by the manufacturer
are for fibers longer than a meter,
while in \mude, we are interested in fiber properties over short lengths ($\leq 30~{\rm cm}$).
Moreover, to our knowledge, data on SciFi timing performance are scarce.

The purpose of this work is to study and compare commercially available blue-emitting
scintillating fibers over short lengths ($< 50~{\rm cm}$)
typical for small to medium-sized SciFi detectors, like beam detectors, small area trackers,
or the \mude SciFi timing detector.
In this work we have studied the properties and performance of different blue-emitting
$250~\mu{\rm m}$ diameter round
double-clad scintillating fibers from Kuraray~\cite{Kuraray,NOL11} and Saint-Gobain~\cite{Gobain},
focusing in particular on their timing properties. 
The scintillating fibers have been excited with ionizing particle beams and radioactive $^{90}{\rm Sr}$ $\beta$ sources.
We hope that the findings presented here will be of interest for the development of new SciFi detectors. 

\section{Characteristics of Scintillating Fibers}
\label{sec:chara}

The role of scintillating fibers is twofold:
i) to convert the deposited ionization energy by the incident particles to optical photons,
ii) to transport the optical signal to the optical readout.
All the fibers discussed in this work are round with a diameter of $250~\mu{\rm m}$.
Fibers with different cross-sections can be produced as well, but most commonly, round fibers are used.
Concerning the diameter, $500~\mu{\rm m}$ and 1~mm diameter fibers are also commonly used,
however at a loss of granularity.
The fibers are made from a polystyrene core with a double cladding structure,
the inner cladding made of polymethyl-methacrylate (PMMA) and the outer cladding of a fluorinated polymer,
with decreasing refractive index with respect to the polystyrene core.
The double cladding is adopted to enhance the scintillation light trapping efficiency,
given that light propagates in the fiber via total internal reflection.
Different organic dyes, 
the {\it activator} luminophores, which convert the deposited radiative energy into UV photons,
and the {\it spectral shifter} luminophores with a large Stoke's shift,
which shift the wavelength of the primary UV photons to visible photons,
are dissolved in the core matrix with different concentrations, as well, depending on the type of fiber.
This is a step by step absorption and re-emission process to visible photons,
which are less absorbed during propagation in the fiber than the primary UV photons.

\begin{table}[!t]
\centering
\begin{tabular}{lcccc}
\hline
fiber type                                          &     SCSF-78     &     SCSF-81     &     NOL-11      &     BCF-12   \\ \hline \hline
cladding thickness [$\%$ fiber radius]  &          \multicolumn{3}{c}{3 / 3}                       &          3 / 1  \\
trapping efficiency [\%]                      &         \multicolumn{3}{c}{5.4}                          &   $\geq 5.6$  \\
numerical aperture                             &         \multicolumn{3}{c}{0.72}                        &   0.74  \\
emission peak [nm]                            &       450         &          437       &        421        &        435    \\
decay time [ns]                                  &        2.8         &          2.4        &           1.3      &          3.2    \\
attenuation length [m]                        &    $> 4.0$        &      $> 3.5$     &      $>2.5$    &         $>2.7$     \\
light yield [ph/MeV]                            &    high           &                       &       high        &  $\sim 8000$  \\
refractive index                                  &    \multicolumn{3}{c}{1.59 / 1.49 / 1.42}           &  1.60 / 1.49 / 1.42  \\
density [${\rm g}/{\rm cm}^2$]          &   \multicolumn{3}{c}{1.05 / 1.19 / 1.43}            &           1.05         \\
core                                                 &  \multicolumn{4}{c}{Polystyrene} \\
inner cladding                                   &  \multicolumn{4}{c}{PMMA} \\
outer cladding                                   &  \multicolumn{4}{c}{Flour-acrylic}  \\
\hline
\end{tabular}
\caption{Properties of different blue-emitting high purity round scintillating fibers from Kuraray~\cite{Kuraray,NOL11} 
and Saint-Gobain~\cite{Gobain}.
The reported values usually are for 1~mm diameter fibers.
Only Saint-Gobain quotes the light yield, whereas Kuraray characterizes SCSF-78 and NOL-11 as "high light yield" fibers.}
\label{tab:fibers}
\end{table}

\begin{figure}[!t]
   \centering
   \includegraphics[width=0.9\textwidth]{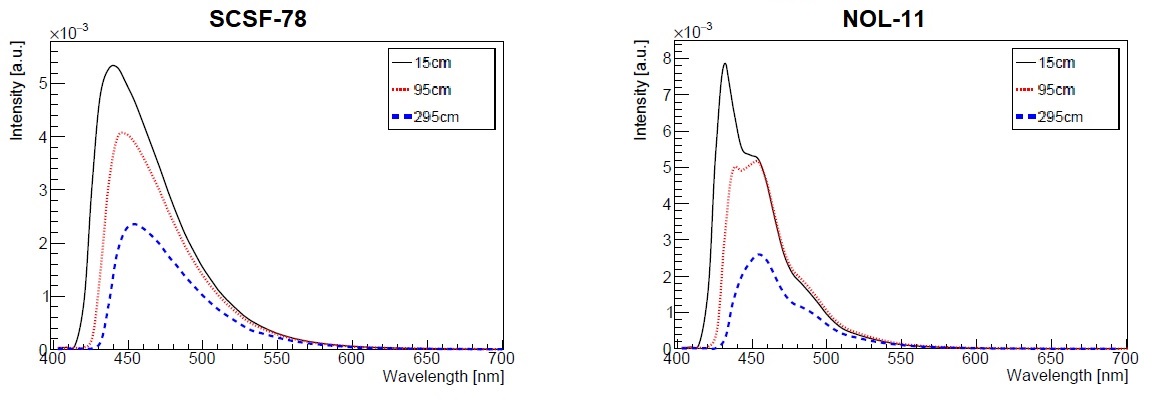}
   \caption{Emission spectra at different distances from the excitation point for the SCSF-78 (left) and NOL-11 (right) fiber
(figure adapted from~\cite{NOL11}).
The spectra are not normalized.}
   \label{fig:fibers}
\end{figure}

Table~\ref{tab:fibers} summarizes the physical properties of various double-clad
round scintillating fibers studied in this work.
The values are as reported by the manufacturer or from the literature,
usually for 1~mm diameter fibers.
The peak emission wavelengths are 450~nm, 437~nm, 421~nm, and 435~nm for
the Kuraray's SCSF-78, SCSF-81, NOL-11, and Saint Gobain's BCF-12 fibers, respectively.
Figure~\ref{fig:fibers} shows an example of emission spectra for the SCSF-78 and NOL-11 fibers
at different distances from the excitation point~\cite{NOL11}.

The performance of a SciFi detector, like the efficiency and the time resolution,
is driven by the signal amplitude $A$ of the optical signal (i.e. the number of detected photons $n_{ph}$),
which can be factorized as:
\begin{equation}
    A(\lambda) = 
          Y_s(\lambda) \cdot \epsilon_\mathrm{trap}(\lambda) \cdot \epsilon_\mathrm{trans}(\lambda) 
          \cdot {\rm PDE}(\lambda) \cdot \Delta E \; ,
\end{equation}
\noindent where $Y_s$ is the ionization light yield,
$\epsilon_\mathrm{trap}$ and $\epsilon_\mathrm{trans}$ are the trapping efficiency and the transport efficiency, respectively,
and PDE is the photo-detection efficiency of the photo-sensor.
All factors depend on the scintillation wavelength $\lambda$,  except for the energy deposit $\Delta E$.

The ionization light yield $Y_s$, expressed in terms of generated visible photons per deposited amount of energy $\Delta E$,
is an intrinsic property of the fiber.
It depends on the fiber's core material, concentration and type of dissolved dyes,
the spectral matching between the {\it activator} and {\it spectral shifter},
and the quantum efficiency of the latter.
Values for $Y_s$ reported in the literature range between 7,000 and 10,000 photons per MeV of deposited energy.

The scintillation photons are emitted isotropically in all directions around the source point,
where the interaction occurred,
but only a fraction, $\epsilon_\mathrm{trap}$, will be captured and transported in the fiber
through total internal reflection,
typically around 5\% of the generated photons per hemisphere.
The transport of the photons through the fiber induces losses due to several effects, 
like disturbances of the internal reflection caused by imperfections in the core material
and defects in the cladding,
light absorption in the fiber due to the overlapping of the emission and absorption bands of the dyes,
which reduces the transparency of the fiber,
and Rayleigh scattering from small density fluctuations in the core,
which can deflect an optical ray which will not be any longer internally reflected.
In addition, a high degree of molecular orientation of the polystyrene core polymers,
which increases the fiber's mechanical strength, reduces its optical transparency. 
This loss during propagation over a distance $d$ is characterized by the light attenuation length $\Lambda(\lambda)$,
which is related to transport efficiency $\epsilon_\mathrm{trans}$ by $\epsilon_\mathrm{trans} = 1 - \exp (d / \Lambda)$.
The attenuation length can be seen as the distance $d$ at which $1/e$ of the initial number of photons survives.
The light intensity $I$ along the fiber as a function of the propagation distance $d$
is usually described in terms of a short and a long component as:
\begin{equation}
    I(\lambda,d) = I_0^\mathrm{short}(\lambda) \cdot  \exp \left(-d / \Lambda^\mathrm{short}(\lambda)\right) + \; ,
                          I_0^\mathrm{long}(\lambda) \cdot  \exp \left(-d / \Lambda^\mathrm{long}(\lambda)\right) \; , 
\label{eq:att}
\end{equation}
\noindent with $\Lambda^\mathrm{short}(\lambda)$ and $\Lambda^\mathrm{long}(\lambda)$
the attenuation lengths for the short and long components, respectively.
This behavior of the intensity $I$ is due to the trajectories of the photons propagating in the fiber,
which can be meridional or non-meridional, and therefore different effects can disturb their propagation.
The meridional trajectories are mainly affected by core effects,
while the non-meridional trajectories by interface effects (losses of light induced by the cladding).
The {\it short} component is predominantly non-meridional and dies off rather quickly after 10 -- 20 cm.
Moreover, the short wavelengths tend to be re-absorbed more strongly than the longer wavelengths,
because of the overlap between the {\it spectral shifter} absorption and re-emission spectra.
Hence, at large distances the emission spectrum is shifted to longer wavelengths
(see Figure~\ref{fig:fibers}).
Most of the fiber's properties reported in the literature are measured over distances of few meters
and only $\Lambda^\mathrm{long}(\lambda)$ is usually reported.

The intrinsic limit on the time resolution of low mass (i.e. thin) scintillation detectors is driven by the statistical processes
involved in the generation of the light signal and fluctuations in the light detection system.
The time resolution of a relatively low light yield SciFi detector strongly depends on the decay time $\tau$ of the {\it spectral shifter}
(the shorter the decay time, the better the achievable time resolution),
but also on the signal amplitude $A$, i.e. the number of detected photons $n_{ph}$.
Assuming that the particle's crossing time is determined by the first detected photon,
the larger the number of detected photons $n_{ph}$, 
the higher the probability that the first detected photon is closer to the fiber's excitation instant,
reducing thus the dependence on $\tau$.
Assuming that the fluorescent states in an organic molecule are formed instantaneously,
like when excited by a UV LED,
the time profile of the light pulse will have a very fast leading edge followed by a simple exponential decay
with time constant $\tau$.
When excited by an ionizing particle, however, the signal formation is a multi-step process
starting with the energy transfer from the radiation source to the an intermediate state, the {\it activator},
followed by the energy transfer from the {\it activator} to the {\it spectral shifter},
and the final emission of light.
If we assume that the population of the intermediate fluorescent states is also exponential,
the overall shape of light pulse of the two-level system is given by~\cite{Birks}:
\begin{equation}
I(t) \sim I_0 \frac{\tau}{\tau - \tau_a} (\exp (-t/\tau) - \exp(-t/\tau_a)) \; ,
\label{eq:birks}
\end{equation}
where $\tau$ is (as before) the time constant describing the decay of the {\it spectral shifter} and
$\tau_a$ describes the population of fluorescent state with $\tau_a$ typically around one~ns.
According to~\cite{Moszynski} the energy transfer from the radiation source to the {\it activator}
is better represented by a Gaussian function.
The time profile of the light pulse is then best described by a convolution of a Gaussian distribution,
which describes the time spread of the light pulse generation process
(and also the fluctuations in the light collection),
with the two step exponential decay function describing the light emission by the {\it spectral shifter}:
\begin{equation}
I(t) \sim \frac{I_0}{\sqrt{2 \pi \sigma^2}} \exp \left( (t-t_0)^2  / 2 \sigma^2 \right) \ast \frac{\tau}{\tau - \tau_a} (\exp (-t/\tau) - \exp(-t/\tau_a)) \; ,
\end{equation}
where $\sigma$ accounts for the light generation process.
The performance of very fast organic scintillators is therefore best described by reporting
the FWHM of the light signal as well as the decay time.

\section{Measurement Setup}
\label{sec:setup2}

\begin{figure}[b!]
    \centering
    \includegraphics[width=0.66\textwidth]{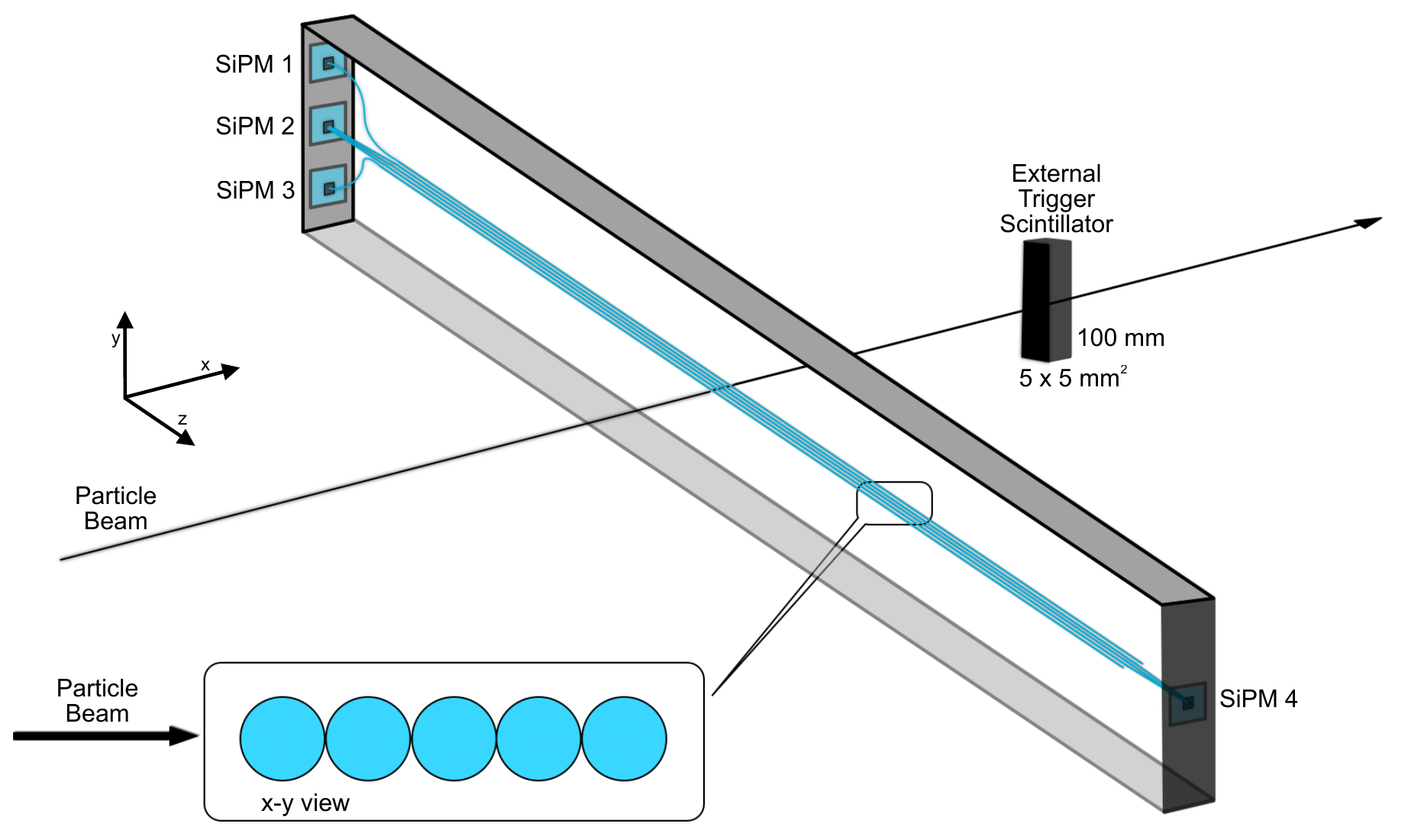}
    \includegraphics[width=0.66\textwidth]{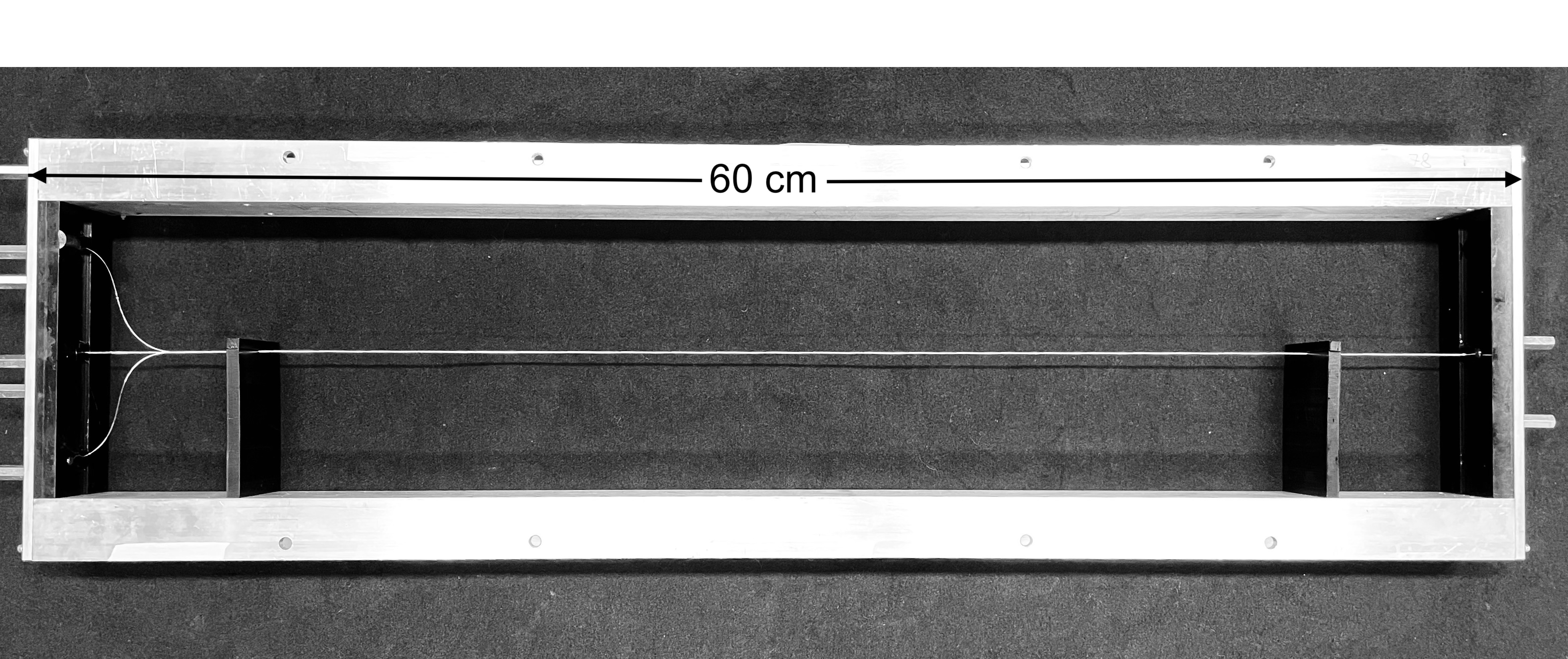}
    \caption{Schematic of the setup showing the mapping of the fibers to the SiPMs (top)
and a picture of the frame holding the fibers (bottom).
The fibers are oriented horizontally w.r.t. the incident beam particles, while the SiPM are positioned vertically in the frame.
The fiber's diameter has been exaggerated for illustration purposes.}
    \label{fig:setup}
\end{figure}

To measure the properties of different scintillating fibers, the fibers have been excited with ionizing particles.
For this purpose,
five $250~\mu{\rm m}$ diameter and 60~cm long fibers have been assembled on top of each other
to form a flat ribbon.
The fiber ribbons have been assembled with a water based ${\rm TiO}_2$ solution 
(Saint-Gobain's reflective paint BC-622A),
which is strong enough to hold the fibers together,
and optically isolates the individual fibers in the ribbon, as well.
The fiber assemblies are placed in frames horizontally w.r.t. incident beam particles, as illustrated in Figure~\ref{fig:setup}.
The top and bottom fibers (i.e. the most upstream and most downstream ones)
function as triggers such to guarantee that charged particles cross all three central fibers.
The central three fibers are grouped together and read out at each end,
while the external fibers are readout on one side only.
All fibers are read out with $1.3~{\rm mm} \times 1.3~{\rm mm}$ SiPMs with $50~\mu{\rm m} \times 50~\mu{\rm m}$
pixels (Hamamatsu device S13360-1350CS~\cite{Hama}).
The fibers are glued in plastic holders, which are used to couple the fibers to the spring loaded SiPMs,
and have been diamond polished.
The Photon Detection Efficiency (PDE) of the SiPMs is quite uniform over a large fraction of the fibers emission spectra
and is close to 40\%,
when the SiPMs are operated at an over voltage (i.e. the voltage above the breakdown) of 3~V.
In addition, a scintillator bar with square cross section of $5~{\rm mm} \times 5~{\rm mm}$ is placed behind the fiber ribbons.
This scintillator bar generates the external trigger and provides the external time reference for the timing measurements.
The trigger scintillator is read out with $3~{\rm mm} \times 3~{\rm mm}$
SiPMs (Hamamatsu device S13360-3050CS~\cite{Hama}) at both ends.
The time resolution of the trigger scintillator has been estimated to be $\sigma_{\rm trigger} \simeq 60~{\rm ps}$.
The signals from the SiPMs were amplified with fast two-stage common emitter transistor amplifiers
(the amplified signals are therefore positive)
with a gain in the order of 35~dB, which provide single photon amplitudes of roughly 100~mV.
The rise time of the amplified signal is around 1.5~ns.
The signals were recorded with the DRS4 waveform digitizer sampling the SiPM signals at 5~GHz
(two synchronized DRS4 Evaluation Boards~\cite{DRS} were used and were read out simultaneously).
For each event the waveforms from all SiPMs were recorded.
The trigger required signals in the outer two fibers and in the square scintillator bar, as well.

The measurements have been performed in the $\pi$M1 beamline at the Paul Scherrer Institute.
The beam was set at a momentum of $161~{\rm MeV}/c$ and contained mainly pions ($> 85\%$).
At this momentum the energy deposited by the pion beam ($\beta \gamma \simeq 1.15$)
is 15\% higher than the energy that a minimum ionizing particles (MIP) would deposit,
and is comparable to the energy loss of highly relativistic particles on the relativistic plateau.
Complementary measurements have been performed with a $^{90}{\rm Sr}$ radioactive source.
A high threshold on the trigger scintillator was applied to select energetic $\beta$ particles ($> 1~{\rm MeV}$),
such that the energy loss is comparable to the energy loss of beam particles.
The timing measurements have been performed with beam only. 

The thickness of the central ribbon, consisting of three fibers, traversed by the ionizing particles ranges from a maximum of three fiber diameters
(i.e. $750~\mu{\rm m}$) down to about half a diameter (i.e. $\sim 60~\mu{\rm m}$). 
The average thickness has been estimated to be around two fiber diameters, i.e. $\sim 500~\mu{\rm m}$,
which corresponds to a 0.1\% of radiation length $X_0$,
taking also into account small misalignments between the fibers and the alignment of the fibers w.r.t. the particle beam.
This thickness corresponds to the thickness of a 4-layer staggered SciFi ribbon of $250~\mu{\rm m}$ diameter fibers.
For different fiber ribbon thicknesses and diameters one should scale up or down the reported results.

\section{Light Yield and Attenuation}
\label{sec:LightYield}

\begin{figure}[t!]
   \centering
   \includegraphics[width=0.95\textwidth]{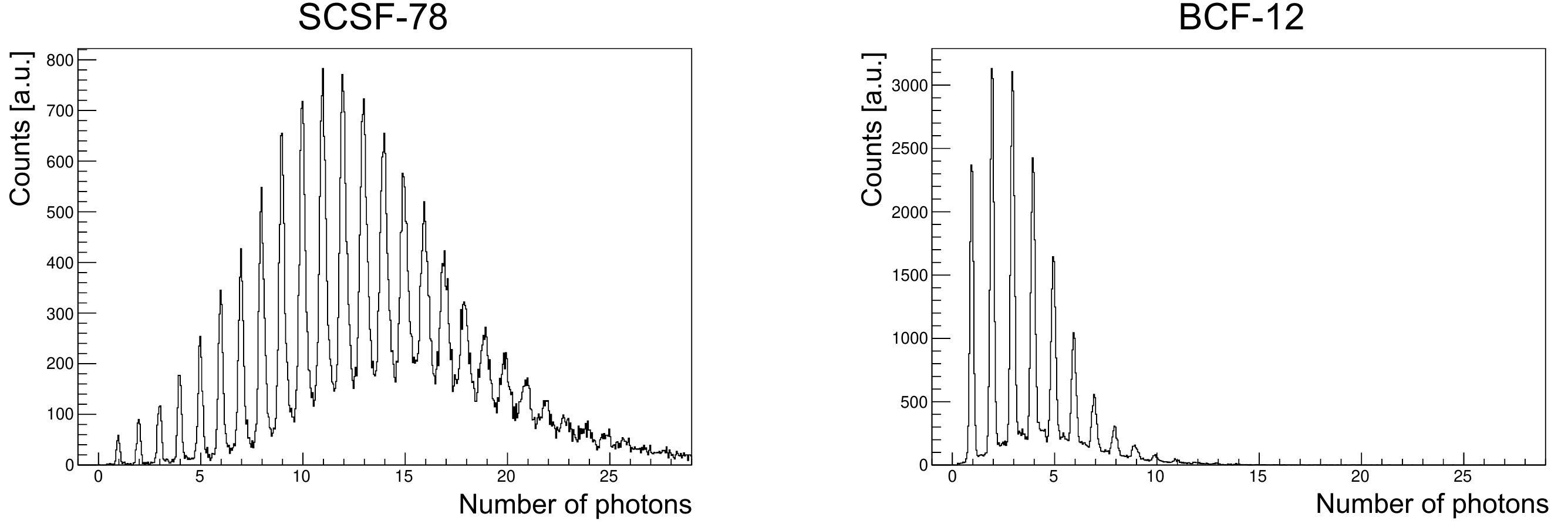}
   \caption{Charge spectrum normalized to the charge generated by a single photon
for a high light yield fiber (SCSF-78, left) and a low light yield fiber (BCF-12, right)
for an average thickness of $\sim 500~\mu{\rm m}$.
The particles cross the fibers in the center of the ribbon, 30~cm from the fiber's ends.}
   \label{fig:light_yield}
\end{figure}

To determine the light yield of each fiber setup
we have evaluated the charge amplitude by integrating the recorded waveforms
after baseline subtraction over a time interval (i.e. the {\it gate}) of 60~ns, starting 5~ns before the start of the signal.
The baseline is evaluated between 50~ns and 25~ns before the start of the signal
for each event and is subtracted on an event by event basis.
Figure \ref{fig:light_yield} shows the charge spectra for a high yield (Kuraray's SCSF-78)
and low light yield (Saint-Gobain's BCF-12) fiber after baseline subtraction.
Thanks to the single photon detection capabilities of the SiPMs,
the charge spectrum can be normalized to the charge generated by one photon,
without knowing the absolute amplification of the SiPM and of the electronics readout chain.
While from these plots one can claim that the efficiency of a SciFi detector using high light yield fibers
can be close to 100\%,
one cannot say the same for low light yield fibers.

\begin{table}[b!]
\centering
\begin{tabular}{l|lll}
\hline
\multicolumn{1}{c|}{Fiber type} & \multicolumn{1}{c}{10~cm} & \multicolumn{1}{c}{30~cm} 
& \multicolumn{1}{c}{50~cm}\\ \hline 
SCSF-78              &   $16.2 \pm 0.9$   &    $13.7 \pm 0.6$   &   $12.5 \pm 0.9$   \\
SCSF-81              &   $  8.1 \pm 0.6$   &    $ 5.8 \pm 0.3$   &   $  5.2 \pm 0.3$   \\
NOL-11               &   $13.0  \pm 1.0$   &   $12.1 \pm 0.9$   &   $10.1 \pm 0.7$     \\
BCF-12               &   $   6.6 \pm 0.4$   &    $ 3.8 \pm 0.1$   &   $  1.5 \pm 0.4$    \\ \hline
\end{tabular}
\caption{Most probable number of detected photons (MPV) for each type of fiber at different distances from the fiber's end.
These values are not corrected for the PDE of the SiPMs, which is around 35\% to 40\%.
The reported errors are form the fits the the charge spectra
and are larger than the differences in the number of detected photons at the two fiber's ends.}
\label{tab:MPV}
\end{table}

\begin{figure}[t!]
   \centering
   \includegraphics[width=0.95\textwidth]{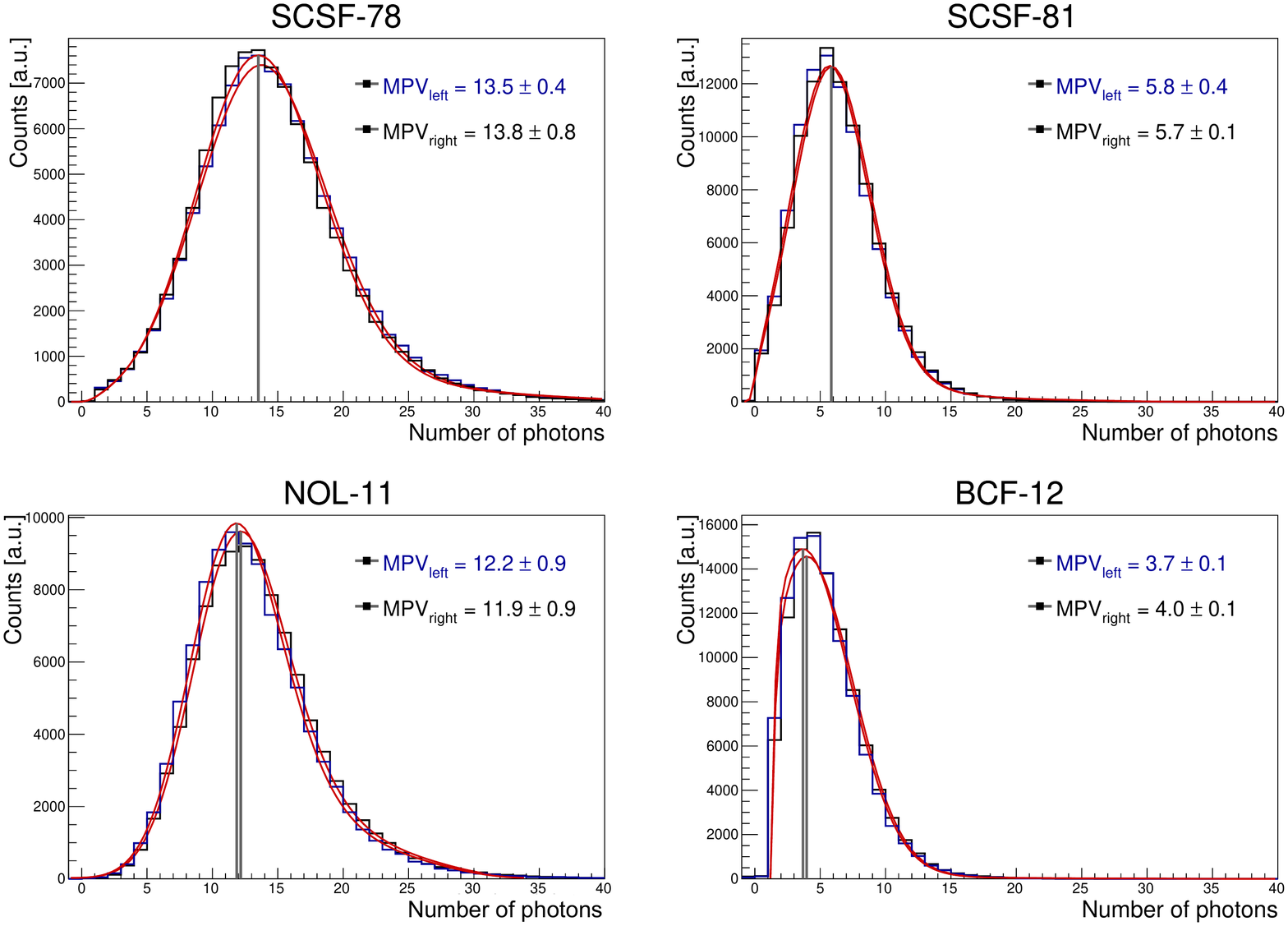}
   \caption{Photon yield for different scintillating fibers measured at both fiber's ends
for an average thickness of $\sim 500~\mu{\rm m}$.
The photon yield is not corrected for the PDE of the SiPMs.
A convolution of a Gaussian and a Landau distribution is fitted to the data
to extract the Most Probable Value (MPV) of the distribution.}
   \label{fig:MPV}
\end{figure}

\begin{figure}[t!]
    \centering
    \includegraphics[width=0.95\textwidth]{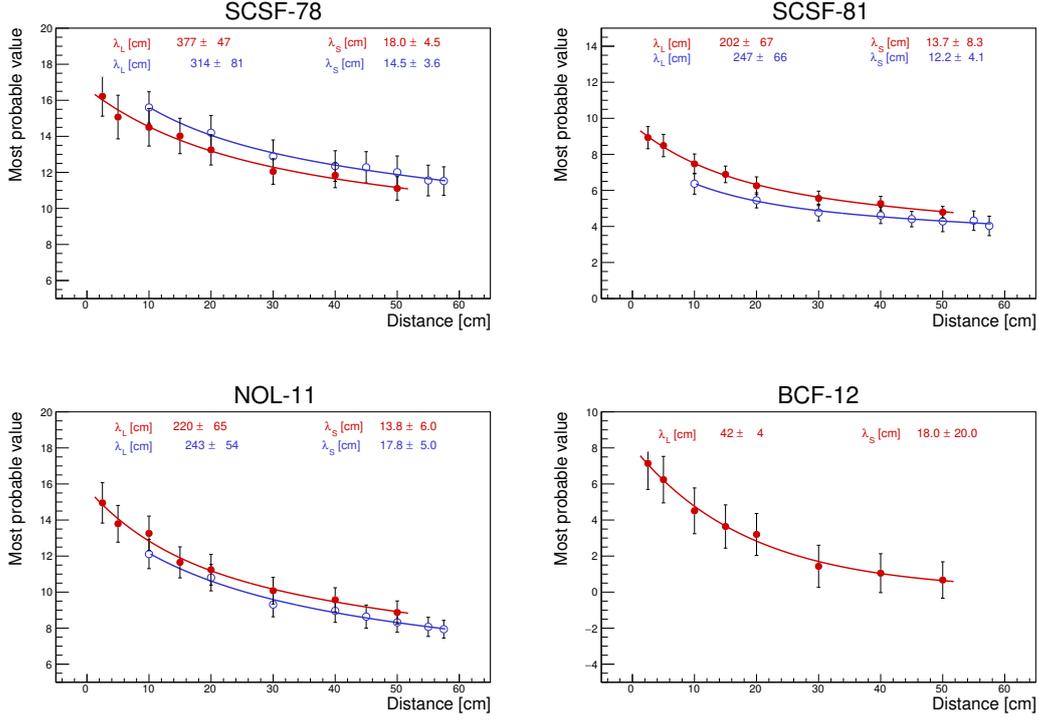}
    \caption{Most probable value MPV as a function of the distance from the fiber's end for different types of fiber.
The full red circles are for measurements from the {\it left} and the open blue circles for the {\it right} fiber's end.
The data points are fitted with a sum of two exponential distributions.
The extracted attenuation lengths are shown in the insets and summarized in Table~\ref{tab:att}.}
    \label{fig:att_d}
\end{figure}

Figure~\ref{fig:MPV} shows the same charge spectra for all 4 fiber types studied in this work.
The normalized charge has been integrated in a region of $\pm 0.5$ around each peak,
and each bin corresponds to the detected number of photons in the event.
The charge spectra have been fitted with a convolution of a Gaussian with a Landau distribution,
which interpolates well the measured spectra.
The Landau distribution describes the fluctuations in the energy deposit in a thin layer of material.
Table~\ref{tab:MPV} summarizes the most probable number of detected photons
(Most Probable Value of the fit, MPV) 
for different fibers and also for different distances from the fiber's end.
The median of the charge distributions is very close to the MPV value. 
The MPV values are not corrected for the PDE of the SiPMs, which is around 35\% to 40\%.
The reported MPV values are obtained by taking the mean of the {\it left} and {\it right} measurements.
The reported errors are from the fits to the charge spectra and are larger than the differences between these measurements.
To verify the reproducibility of the measurements,
the light yield at both ribbon's ends are compared in Figure~\ref{fig:MPV} and~\ref{fig:att_d}.
Some measurements have been repeated using different SiPMs of the same type
and different fiber assemblies using the same fiber, as well.
All measurements were very close to each other. 
As can be appreciated from Figure~\ref{fig:MPV} the measured values differ by less than one detected photon.

Here we report only the relative light yields of the fibers using the same {\it normalization},
which allow for a direct comparison of different scintillating fibers.
Estimating the intrinsic ionization light yield $Y_s$ of a scintillating fiber, on the contrary, is a challenging task,
since it requires the precise knowledge of the photo-sensor PDE, the efficiency of the optical coupling,
the propagation of light in the fiber (trapping efficiency and attenuation), and other parameters, as well.

\begin{table}[!b]
\centering
\begin{tabular}{l|ll}
\hline
\multicolumn{1}{c|}{Fiber type} & 
\multicolumn{1}{c}{$\Lambda^\mathrm{short}$~[cm]} & \multicolumn{1}{c}{$\Lambda^\mathrm{long}$~[cm]} \\ \hline 
SCSF-78              &  $16.4 \pm  4.1$    &  $337 \pm  62$   \\
SCSF-81              &  $13.2 \pm  6.3$    &  $224 \pm  66$   \\
NOL-11               &  $15.6 \pm  5.4$    &  $230 \pm  59$  \\
BCF-12                &  $18.2 \pm 20 $    &  $  42 \pm  4  $ \\ \hline
\end{tabular}
\caption{Short and long components of the attenuation length for different fibers.}
\label{tab:att}
\end{table}

To study the light attenuation we excited the fibers at different distances from the fiber's end (from 5~cm to 55~cm)
and measured the charge spectra for each point. 
As discussed earlier, at this lengths, the absorption of light is controlled by both the short component
$\Lambda^\mathrm{short}$ and the long component $\Lambda^\mathrm{long}$.
Figure~\ref{fig:att_d} shows the variation in the detected number of photons (MPV) as a function of the distance
from the fiber's end.
Measurements for both fiber ends are shown.
The small differences are due to the optical couplings of the fibers to the SiPMs, the PDE of each SiPM,
non-uniformities in the fibers,
misalignments in the assembly of the fiber ribbons, etc.
From these differences one can also asses the reproducibility of the measurements.
The measurements are fitted with a sum of two exponential distributions (Equation~\ref{eq:att}).
The extracted attenuation lengths are summarized in Table~\ref{tab:att}.
The reported values are the weighted sum of the {\it left} and {\it right} measurements,
except for the BCF-12 fiber,
for which only the measurements from the {\it left} side are reported.
The reason for not fitting the {\it left} and {\it right} side measurements together
is that the two measurements cover different ranges and are affected by small differences in the detection efficiency,
which would over-constrain the fit by the measurements at the extremities.
The reported errors are large, because of the relatively short lever arm and the fact that 
the attenuation lengths have been determined in a region where the {\it short} and {\it long}
components contribute.
$\Lambda_{\rm short}$ is around 10 -- 20~cm as anticipated earlier,
while $\Lambda_{\rm long}$ is in reasonable agreement with the values reported in the literature
(see Table~\ref{tab:fibers}) except for the BCF-12 fiber.
Note, however, that Table~\ref{tab:fibers} reports values for 1~mm diameter fibers
and this could explain the behavior of the BCF-12 fiber observed here.

\section{Timing Performance}

\begin{figure}[t!] 
   \centering
   \includegraphics[width=0.95\textwidth]{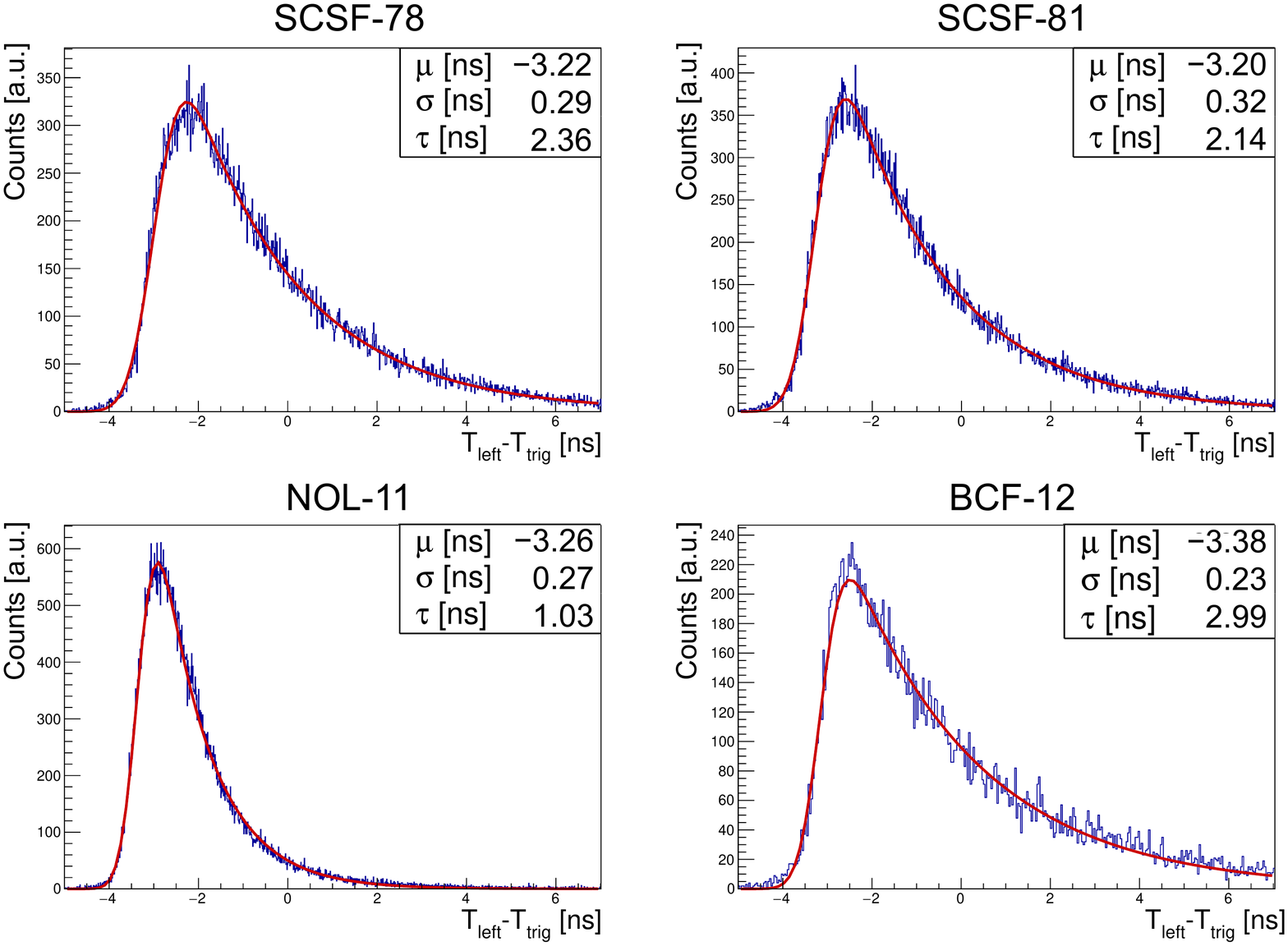}
   \caption{Arrival time of the first detected photon
(i.e. the time difference $T_{\rm fiber} - T_{\rm trigger}$ between the first detected photon and the external time reference)
for different types of fiber.
The decay time of the fiber ribbon is extracted by fitting an EMG distribution to the data.}
\label{fig:TDecay}
\end{figure}

\subsection{Scintillation Light Decay Time}
\label{sec:Tdecay}

The decay time of the scintillation light can be obtained from the time difference between
the detection of the first scintillation photon ($T_{\rm fiber}$)
and the external timing reference given by the trigger scintillator ($T_{\rm trigger}$).
The time resolution of the trigger scintillator is $\sigma_{trigger} \simeq 60~{\rm ps}$
and does not affect significantly the shape of the light pulse.
The arrival time of the first scintillation photon is determined from the analysis of the recorded waveforms.
It is extracted by interpolating the rising edge of the signal with a straight line,
and extrapolating the interpolated line to the baseline of the waveform,
after correcting for baseline fluctuations.
The interpolation is performed on 4 waveform samples on the rising edge of the waveform,
the first sample is below the single photon half amplitude, the next three samples above.
At 5~GHz sampling, the samples are spaced by $\sim 200~{\rm ps}$ for a time base of $\sim 600~{\rm ps}$.
The time walk of this algorithm has been estimated to be below 10~ps
(see the discussion following Figure~\ref{fig:MeanTn}).
Different algorithms have been tried, as well,
like a leading edge discriminator with a very low threshold of half photon amplitude,
or a constant fraction algorithm with variable ratios. 

Figure~\ref{fig:TDecay} shows the arrival time of the first detected photon
(i.e. the time difference $T_{\rm fiber} - T_{\rm trigger}$ between the first detected photon and the external time reference)
distribution for different fibers.
The distributions are fitted with a convolution of an exponential and a Gaussian distribution,
the so called exponentially modified Gaussian distribution (EMG) or exGaussian distribution: 
\begin{equation}
F(t, \mu, \sigma, \tau) = A \frac{1}{2\tau} \exp \left( \frac{\mu - t}{\tau} + \frac{\sigma^2}{2\tau^2} \right)
\left[ 1 - {\rm erf} \left( \frac{1}{\sqrt 2} \left( \frac{\mu - t}{\sigma} + \frac{\sigma}{\tau} \right) \right)  \right] \; ,
\label{eq:EMG}
\end{equation}
where $\tau$ is the scintillation light decay time and $\sigma$ accounts for the time spread of the light collection
(which includes also the time jitter of the external time reference,
the transit time jitter of a single photon in the SiPMs of $\sim 200~{\rm ps}$,
the electronics response, etc.).
erf is the error function, $t = T_{\rm fiber} - T_{\rm trigger}$, and $A$ is a normalization factor.

\begin{figure}[t!] 
   \centering
   \includegraphics[width=\textwidth]{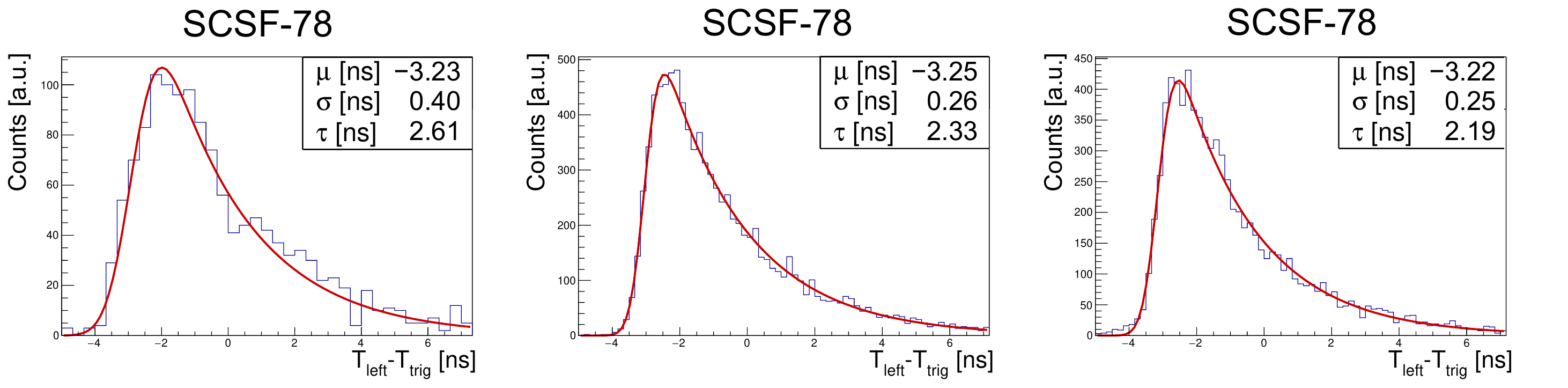}
   \caption{Decay time distribution for
$n_{ph} \leq 5$ (left), $10 < n_{ph} \leq 15$ (center), and $n_{ph} > 20$ (right) for the SCSF-78 fiber
fitted with the EMG distribution.}
   \label{fig:TDecay2}
\end{figure}

\begin{figure}[b!] 
   \centering
   \includegraphics[width=0.95\textwidth]{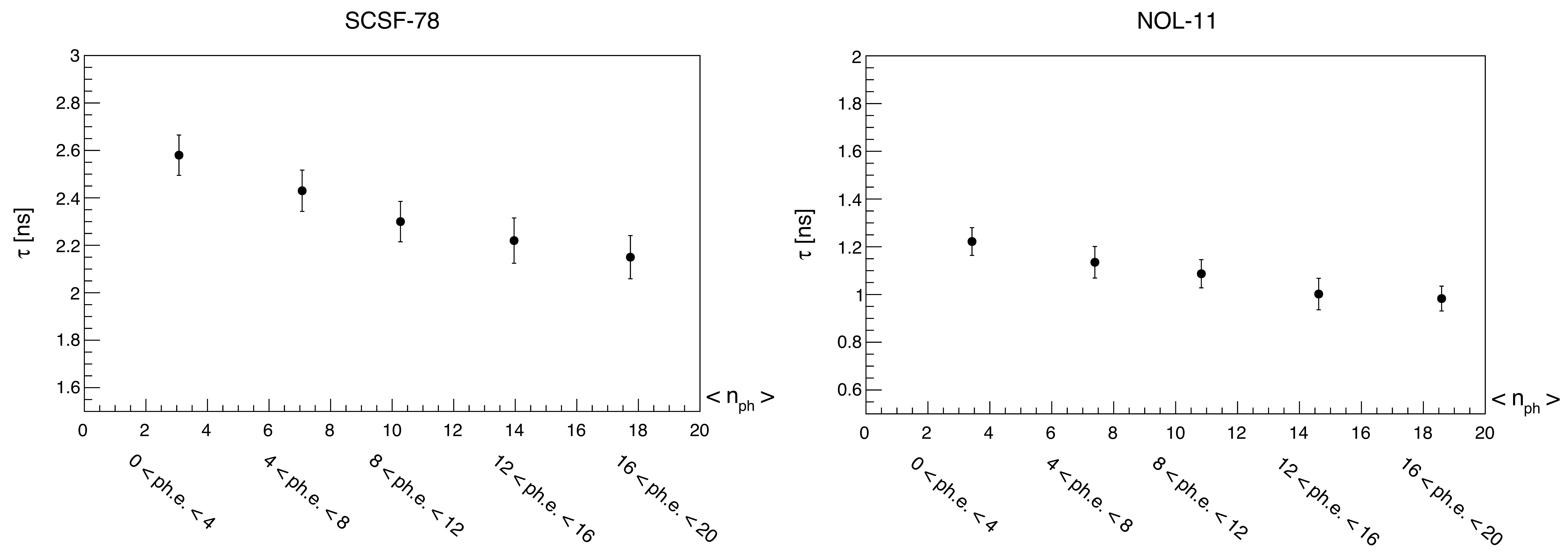}
   \caption{Decay time of the scintillation light
as function of the number of detected photons $n_{ph}$ for the SCSF-78 fiber (left)
and NOL-11 fiber (right).}
   \label{fig:TDecay3}
\end{figure}

Since the photons are generated randomly in the de-excitation of several {spectral shifters}
the observed time arrival of the first detected photon distribution is an inhomogeneous Poisson point-process
in which the random sequence of the photon arrival times can be described by its mean time $\tau$.
The extracted values for $\tau$ are smaller than those reported in Table~\ref{tab:fibers},
but follow the same {\it hierarchy}.
Strictly speaking, the extracted decay time $\tau$ does not measure the decay time of the {\it spectral shifter},
since the signal formation is a multi step process and
many molecules are excited at the same time by the ionizing particle,
while the first arriving detected photon determines $T_{\rm fiber}$.
As it can be seen in Figures~\ref{fig:TDecay2} and~\ref{fig:TDecay3},
$\tau$ depends on the number of detected photons $n_{ph}$.
Figure~\ref{fig:TDecay2} shows the decay time distribution for different numbers of detected photons $n_{ph}$
for the SCSF-78 fiber.
As the number of photons $n_{ph}$ increases, the distribution shrinks and the tail on the right of the peak dies off very slowly.

We have also studied the decay time for different excitation points from the fiber's end
and did not observe appreciable variations.

Incidentally, Figure~\ref{fig:TDecay} gives also an appreciation of the time resolution achievable
with single ended readout,
which is dominated by the decay time for low photon statistics.
The variance $\sigma_t$ of the EMG distribution is given by
\begin{equation}
   \sigma_t = \sqrt{\sigma^2 + \tau^2} \; ,
\end{equation} 
\noindent where $\sigma$ is the intrinsic resolution of the system.
$\sigma_t$ is not the best choice for quantifying the time resolution of the fiber
given the asymmetric nature of the timing distribution.
One possibility for quoting the time resolution achievable, which takes into account the nature of the light generation process,
is to give asymmetric errors by taking the widths of the distribution for the left $\sigma_{\rm left} = \sigma$ and
right $\sigma_{\rm right} = \sqrt{\sigma^2 + \tau^2}$ sides of the peak separately.

\subsection{Time Difference $\Delta T = T_{\rm left} - T_{\rm right}$}

\begin{figure}[t!]
   \centering
   \includegraphics[width=0.95\textwidth]{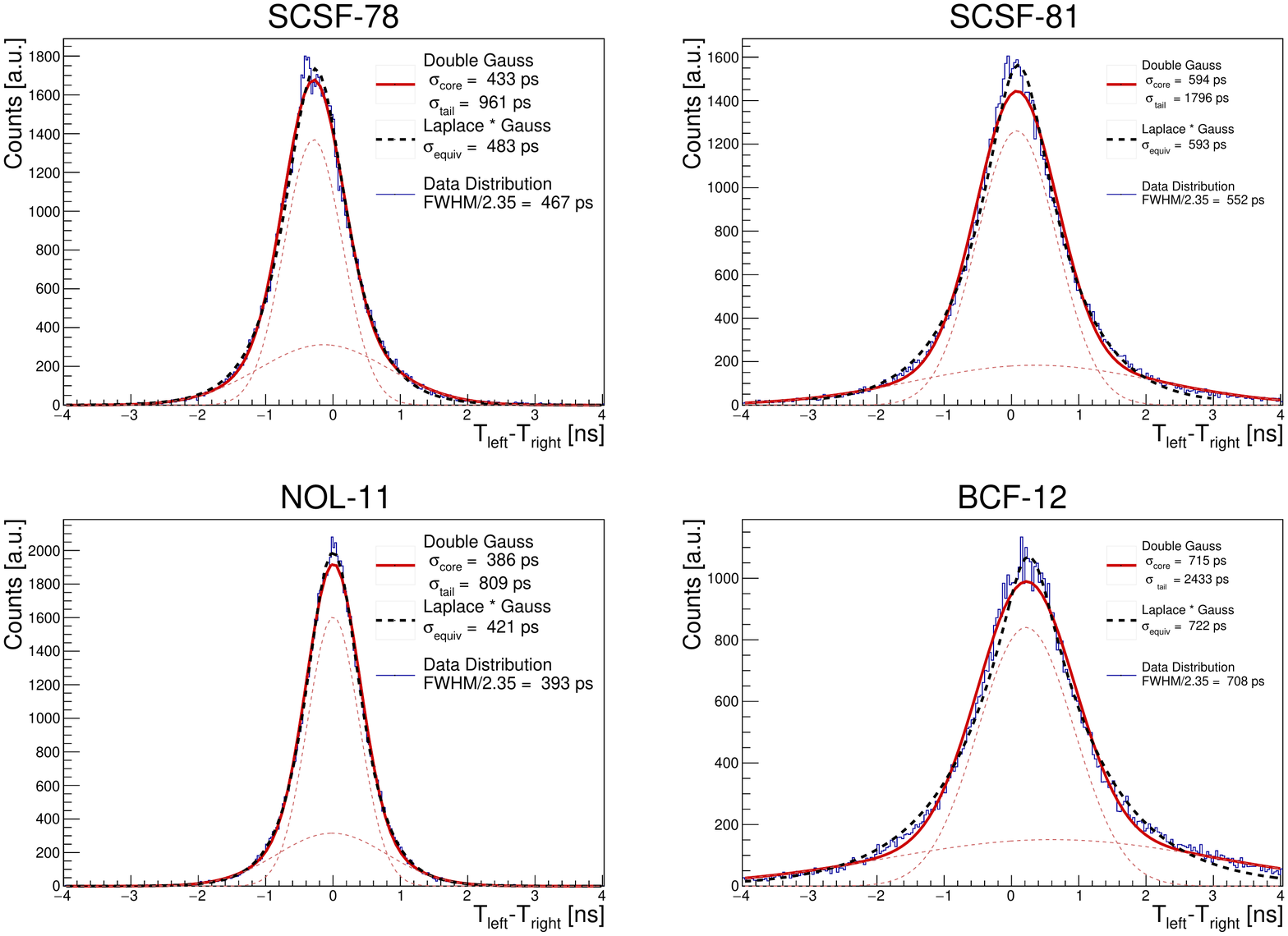}
   \caption{Time-difference $\Delta T$ for different types of fiber.
The $\Delta T$ distributions are fitted with a sum of two Gaussian distributions centered around a common mean value (red lines),
and with a convolution of a Laplace distribution and a Gaussian distribution (black line).}
   \label{fig:DeltaT}
\end{figure}

The time resolution can be much improved by reading out the fibers at both ends and combining both
time measurements.
To begin, we have estimated the time-difference resolution of the fibers
by studying the time difference $\Delta T = T_{\rm left} - T_{\rm right}$,
where $T_{\rm left}$ and $T_{\rm right}$ are the time measurements for the {\it left} and {\it right} fiber ends.
$\Delta T$ is self-contained in the sense that it can be formed without an external time reference.
$\Delta T$, however, cannot be used to determine the crossing time of a particle.
The $\Delta T$ distribution is symmetric around the peak value,
since the fluctuations in the time measurements add/subtract symmetrically for the {\it left} and {\it right} fiber ends.
Figure~\ref{fig:DeltaT} shows the $\Delta T$ distributions for different fiber types
for beam particles crossing the fiber ribbons in the center (i.e. 30~cm from the fiber's ends).
The $\Delta T$ distributions are not necessarily centered around 0, for instance because of different cable lengths. 
The large tails, which extend symmetrically around the peak, are driven by the fiber's decay time.

The $\Delta T$ distributions are fitted
with the sum of two Gaussian distributions centered around a common mean value:
the first Gaussian, which describes the core of the distribution, can be interpreted as due to fluctuations in the light collection,
while the second Gaussian describes the tails, which are driven by the fiber's decay time.
Around 80\% of events fall under the first Gaussian, while the remaining 20\% under the second.
As an indication of the time-difference resolution $\sigma_{\Delta T}$ we quote the FWHM/2.355 of the $\Delta T$ distribution,
which is close to the width of the first Gaussian.
The measured values range between 400 and 700~ps and
the best resolution of $\sim 400~{\rm ps}$ is obtained with the NOL-11 fiber,
which has the shortest decay time.
To be noted that the SCSF-81 fiber, which has a slightly shorter decay time compared to the SCSF-78 fiber,
generates significantly larger tails.
This is mainly due to the low light yield of the SCSF-81 fiber.
The poor performance of the BCF-12 fiber is due also to the low light yield of this fiber.

\begin{figure}[t!]
   \centering
   \includegraphics[width=0.87\textwidth]{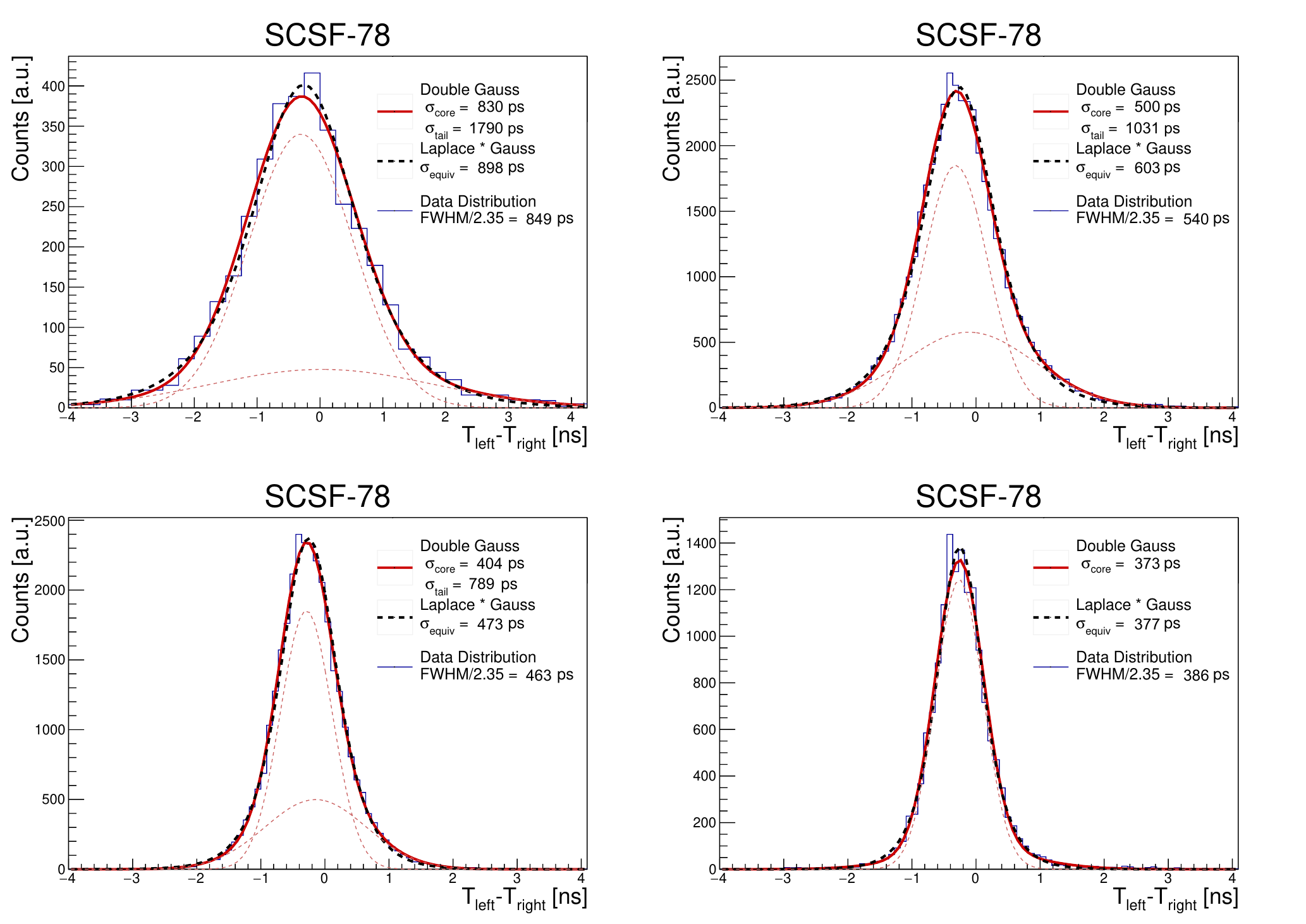}
   \includegraphics[width=0.87\textwidth]{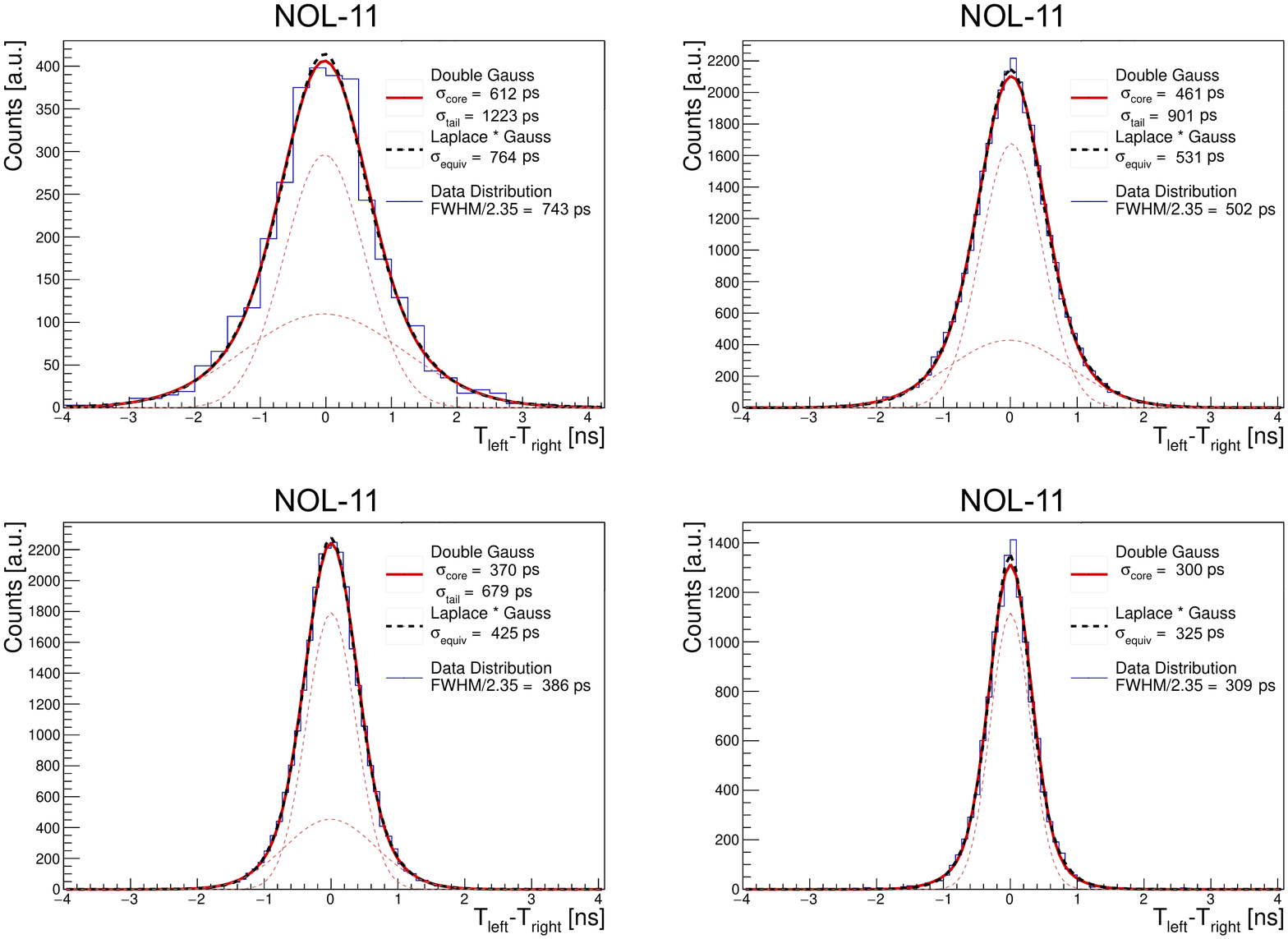}
   \caption{Time-difference $\Delta T$ distribution for different number of detected photons
$\Sigma n_{ph} \leq 10$ (top left), $10 < \Sigma n_{ph} \leq 22$ (top right), 
$22 < \Sigma n_{ph} \leq 32$ (bottom left), and $\Sigma n_{ph} > 32$ (bottom right)
for the SCSF-78 (top four) and the NOL-11 (bottom four) fibers.
As $\Sigma n_{ph}$ increases, the tails disappear and the distribution tends toward a Gaussian.
For $\Sigma n_{ph} > 30$ the $\Delta T$ distribution can be described by a single Gaussian.
The fitted lines are as in Figure~\ref{fig:DeltaT}.} 
   \label{fig:DeltaTn}
\end{figure}

Alternatively, the $\Delta T$ distribution can be modeled with a symmetric exponential function such as the Laplace distribution,
which is a direct consequence of the Poissonian nature of the scintillation process,
smeared with a Gaussian distribution to take into account the time spread of the light collection:
\begin{equation}
f(t) = A \cdot \exp \left( -|t-t_0| / \tau \right) \ast \frac{1}{\sqrt{2 \pi \sigma^2}} \exp \left( (t-t_0)^2 / 2 \sigma^2 \right) \; .
\end{equation}
This convolution models rather well the $\Delta T$ distribution, in particular the tails (black line in Figure~\ref{fig:DeltaT}).
For the spread of the $\Delta T$ distribution we quote the FWHM/2.355 of the convolution.
As it can be deduced from Figure~\ref{fig:DeltaTn}, the time constant $\tau$ scales with $n_{ph}$.
 
\begin{figure}[t!]
   \centering
   \includegraphics[width=0.95\textwidth]{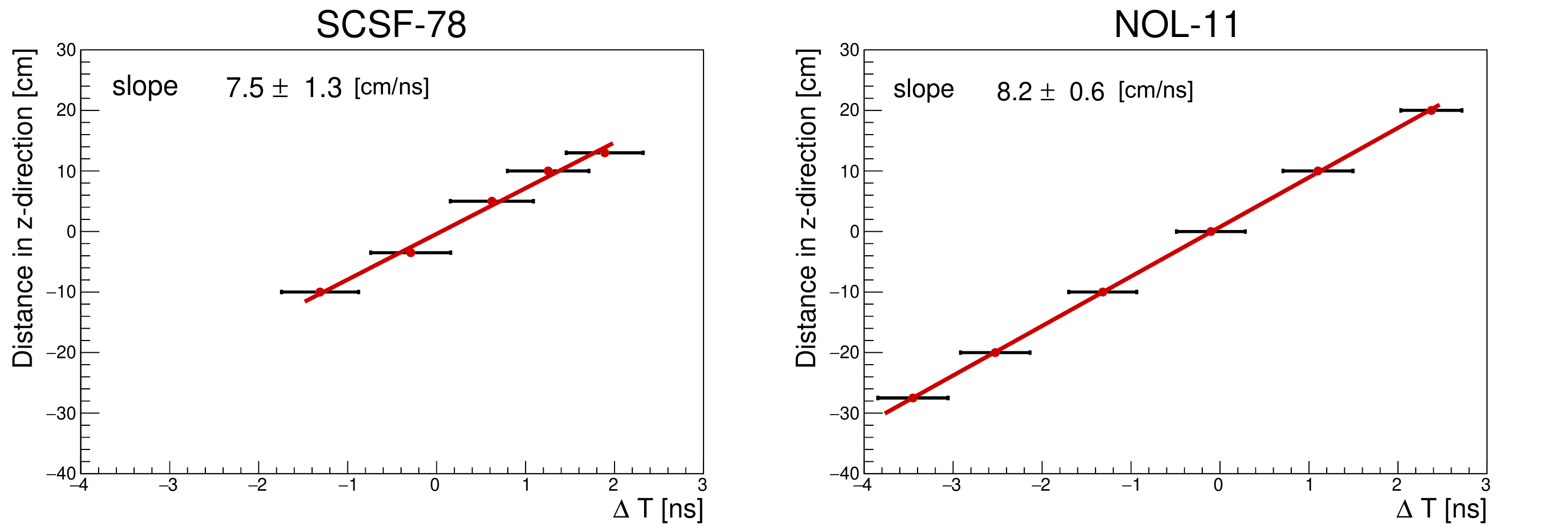}
   \caption{$\Delta T$ as a function of the position for the SCSF-78 (left) and NOL-11 (right) fibers.
The widths of the $\Delta T$ distributions do not change significantly with position.
The data points are interpolated with a straight line.
Taking into account that the difference in the traveled distance by the photons is twice the displacement,
from the slope one can derive the speed of light propagation in the fibers,
which is found to be $v_{\rm fiber} \sim 0.5 \times c$.} 
   \label{fig:DeltaTz}
\end{figure}

\begin{figure}[b!]
   \centering
   \includegraphics[width=0.95\textwidth]{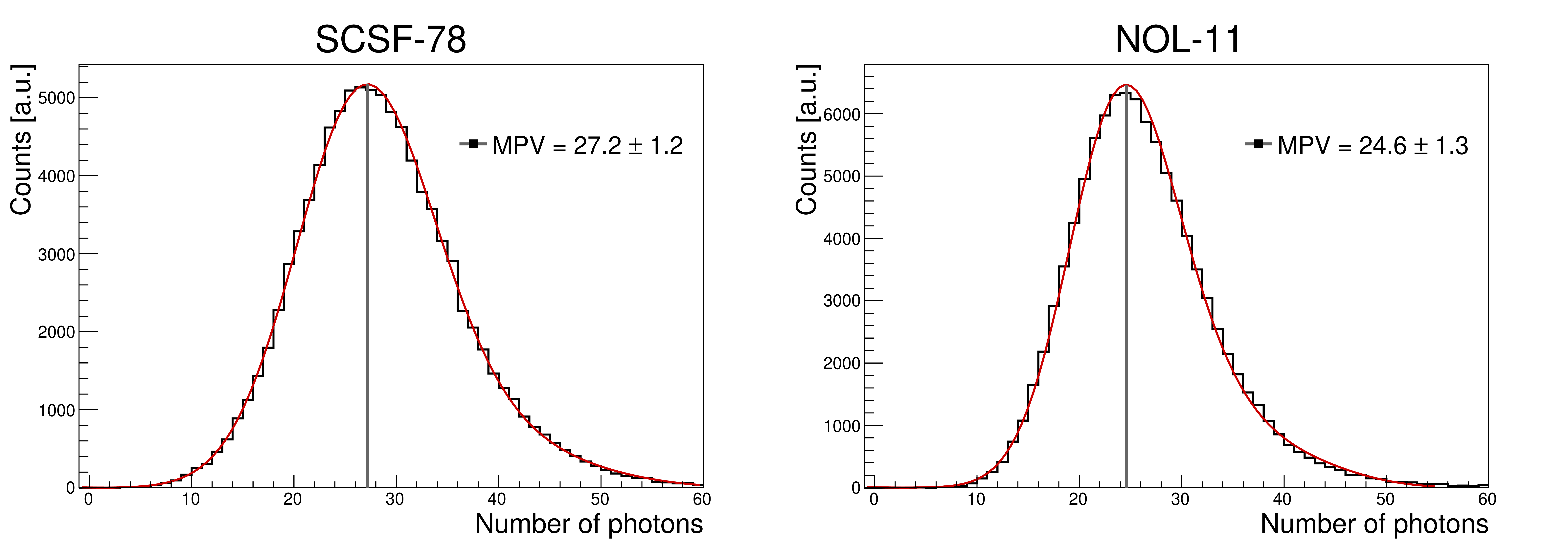}
   \caption{Total number of detected photons $\Sigma n_{ph}$
obtained by summing $n_{ph}$ at both fiber ends for the SCSF-78 (left) and NOL-11 (right) fibers.
The distributions are fitted with a convolution of a Gaussian and a Landau distribution
(not corrected for the PDE of the SiPMs).} 
   \label{fig:MPVsum}
\end{figure}

Figure~\ref{fig:DeltaTz} shows $\Delta T$ for different positions along the fiber.
Since the distance to the photo-sensor changes,
the peak of the $\Delta T$ distribution moves accordingly.
We have found that the width of the $\Delta T$ distribution remains almost constant,
indicating that the time-difference resolution does not depend on the impact position.
The data points can be interpolated with a straight line
supporting the idea that the light propagates uniformly in both directions.
Within our resolution, we do not observe edge effects as we approach the end of the fiber. 
Taking into account that the difference in the traveled distance by the photons is twice the displacement,
from the slope one can derive the speed of light propagation in the fibers $v_{\rm fiber}$.
We find that the photons propagate at the speed,
which is roughly half of the speed of light in vacuum, i.e. $v_{\rm fiber} \sim 0.5 \times c$.
$v_{\rm fiber}$ is significantly slower
compared to the speed that one would derive from the refractive index $n$ of the fiber core material, i.e. $c/n$.
This can be understood by the fact that during propagation the photons are internally reflected from the cladding many times and 
therefore travel over a longer distance. 
From $\Delta T$ one can also determine the impact position $z$ along the fiber 
$z = \frac{1}{2} L + \Delta T \times \frac{1}{2} v_{\rm fiber}$ with
a position resolution of $\sigma_z = \sigma_{\Delta T} \times \frac{1}{2} v_{\rm fiber} \sim 3.5~{\rm cm}$,
where $L$ is the length of the fiber.
While modest, this spatial resolution can be useful in some applications.

To study the $\Delta T$ distribution dependence on the number of detected photons
we first sum the number of detected photons at both fiber ends $\Sigma n_{ph}$
(see Figure~\ref{fig:MPVsum}).
Figure~\ref{fig:DeltaTn} shows the $\Delta T$ distribution for different intervals of $\Sigma n_{ph}$.
As $\Sigma n_{ph}$ increases, the $\Delta T$ distribution shrinks and the tails die off,
and the $\Delta T$ distributions tends toward a Gaussian distribution;
in other words the time-difference resolution $\sigma_{\Delta T}$ improves.
For $\Sigma n_{ph} > 30$ the $\Delta T$ distribution is approaching a Gaussian distribution,
since the tails have been fully absorbed.
From this analysis, it is evident that, in order to achieve the best timing,
it is not only important to minimize the scintillation light decay time $\tau$,
but also to maximize the number of detected photons $n_{ph}$.

\subsection{Mean Time {\it MT}}

\begin{figure}[t!]
   \centering
   \includegraphics[width=0.95\textwidth]{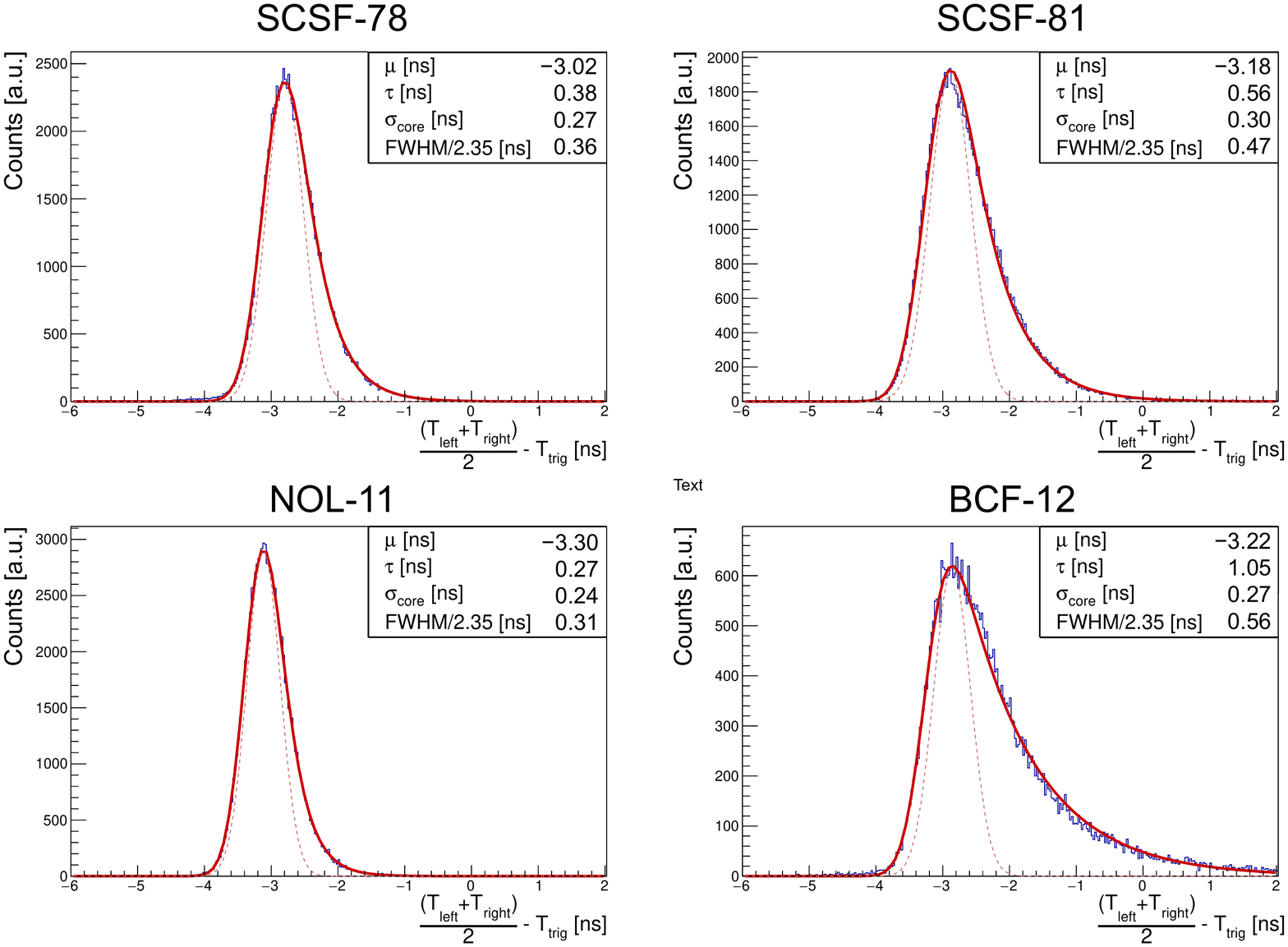}
   \caption{Mean Time $MT$ for different types of fiber.
The $MT$ distributions are skewed with a tail extending on the right of the peaks.
The $MT$ distributions are fitted with the EMG distribution.
Also superimposed is a Gaussian distribution with width $\sigma_{\rm core}$
(dashed line).}
   \label{fig:MeanT}
\end{figure}

The usual approach to measure time, when reading out a scintillator at both ends,
is to form the so called Mean Time $MT$,
defined as $MT = \frac{1}{2}(T_{\rm left} + T_{\rm right}) - T_{\rm trigger}$,
where $T_{\rm trigger}$ represents an arbitrary reference time,
which could also come from a second timing measurement, like in a time of flight measurement,
or from the system clock.
In the study of the $MT$ distribution, $T_{\rm trigger}$ can be considered as an additive constant,
provided that $\sigma_{\rm trigger}$ is small compared to the width of the $MT$ distribution.
The principal reason for forming the mean time is that $MT$ does not depend, to a good degree of accuracy,
on the hit position in the scintillator.
Hence one does not need to take into account the light propagation time in the scintillator
and to correct for it.
Contrary to $\Delta T$, the external time reference,
which can be given by an external trigger or system clock, does not cancel in the sum.
By construction one would expect that the width of the $MT$ distribution is half the width of the $\Delta T$ distribution,
i.e. $\sigma_{MT} = 1/2 \sigma_{\Delta T}$.
In general this assumption holds, provided that the time measurements are normally distributed,
as would be the case with a thick scintillator.
The $MT$ distributions for the different types of fiber are shown in Figure~\ref{fig:MeanT}.
These measurements have been taken with the beam crossing the fibers in the middle.
As it can be observed, the $MT$ distribution is not symmetric w.r.t. the peak with a small tail extending to the right of the peak,
which is driven by fiber's decay time.
The $MT$ distributions in Figure~\ref{fig:MeanT} are well described with the EMG distribution
like the decay times in Figure~\ref{fig:TDecay}.
Since the $MT$ distributions are not too skewed, especially for the NOL-11 and SCSF-78 fibers,
we can take $\sigma_{\rm core}$ of the $MT$ distribution as indication for the time resolution $\sigma_{MT}$.
Otherwise one could report asymmetric errors for the left and right sides of the peak, as mentioned earlier.
Neglecting the jitter of the external time reference $\sigma_{\rm trigger} \simeq 60~{\rm ps}$,
the time resolution $\sigma_{MT}$ achievable with the NOL-11 fiber is $\simeq 250~{\rm ps}$,
while it is $\simeq 275~{\rm ps}$ for the SCSF-78 fiber.
This is about 30\% larger than $1/2 \sigma_{\Delta T}$ (but significantly smaller than $\sigma_{\Delta T}$)
and it is driven by the decay tail of the scintillation light.

\begin{figure}[t!]
   \centering
  \includegraphics[width=0.95\textwidth]{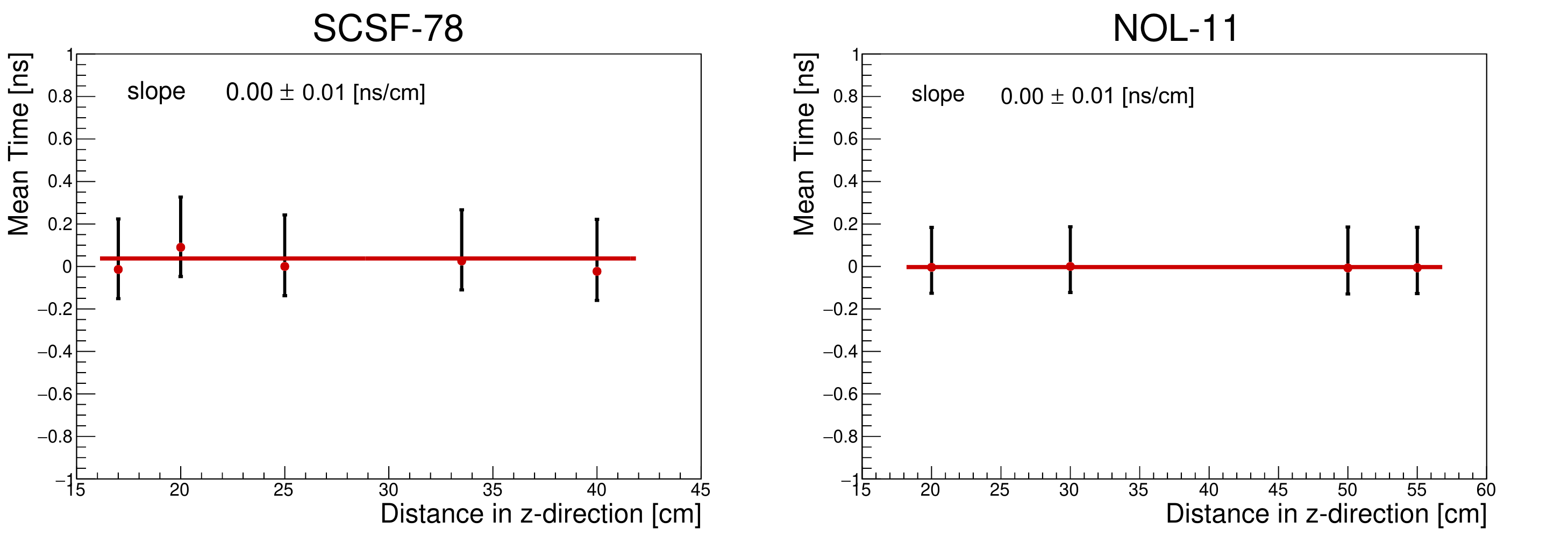}
   \caption{Mean time $MT$ distribution peak position as a function of the distance from one fiber's end
for SCSF-78 (left) and NOL-11 (right) fibers.
The data points are interpolated with a horizontal line showing that there is no appreciable position dependence.
The widths of the $MT$ distributions do not change significantly, as well.} 
   \label{fig:MeanTz}
\end{figure}

\begin{figure}[t!]
   \centering
   \includegraphics[width=0.87\textwidth]{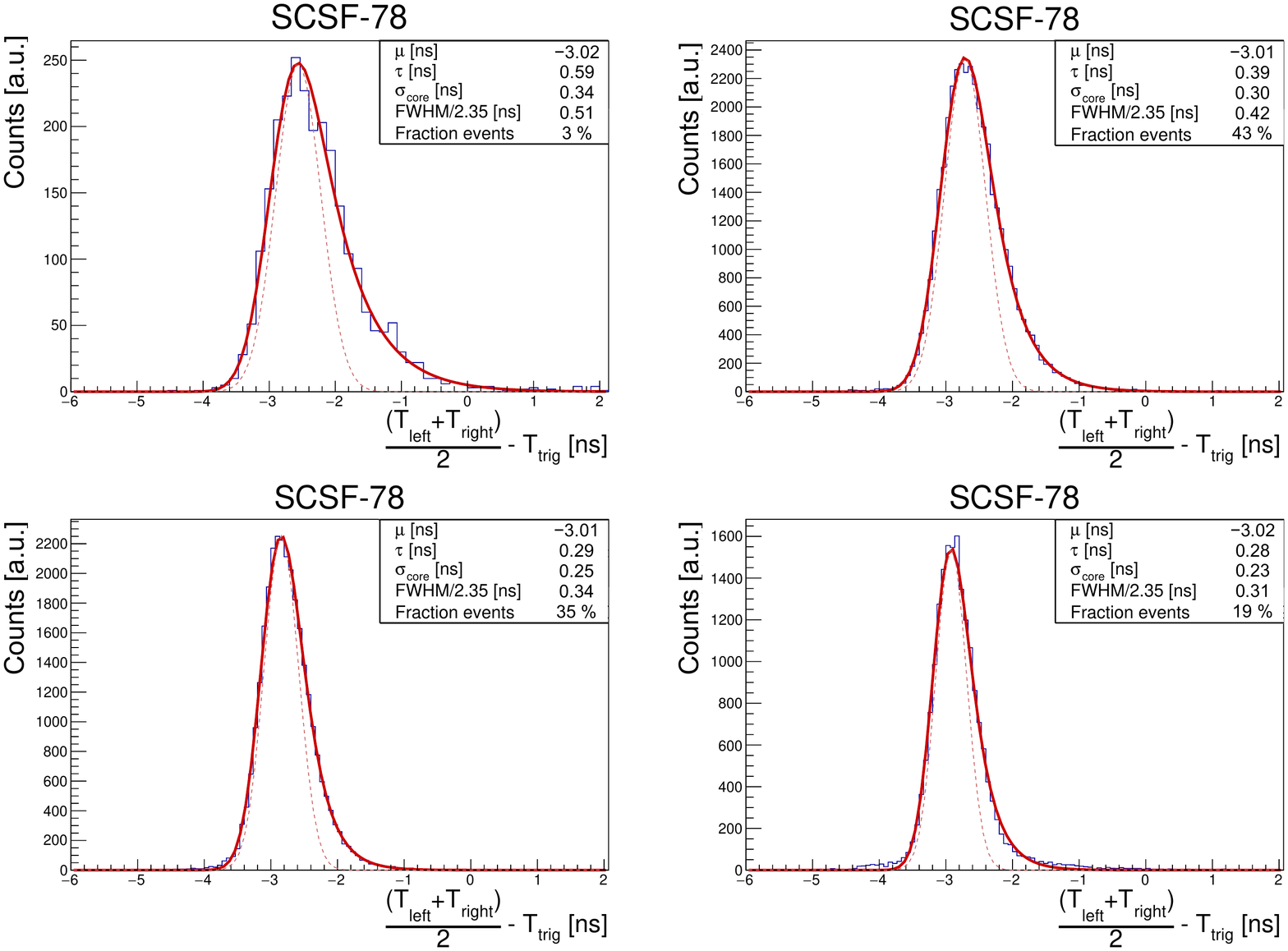}
   \includegraphics[width=0.87\textwidth]{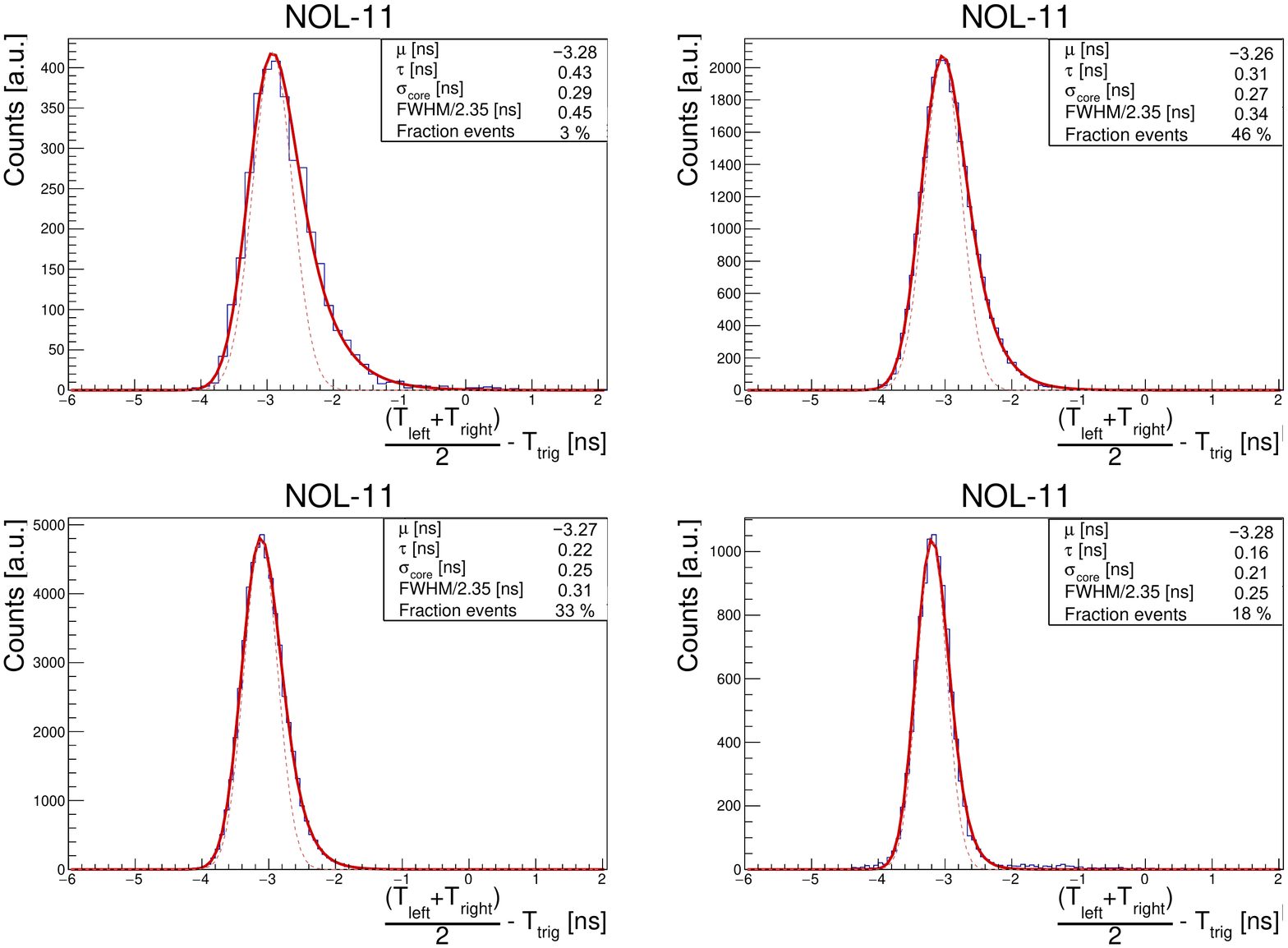}
   \caption{$MT$ distribution for different number of detected photons
$\Sigma n_{ph} \leq 10$ (top left), $10 < \Sigma n_{ph} \leq 22$ (topright), 
$22 < \Sigma n_{ph} \leq 32$ (bottom left), and $\Sigma n_{ph} > 32$ (bottom right)
for the SCSF-78 (top four) and the NOL-11 (bottom four) fibers.
As $\Sigma n_{ph}$ increases, the tail dies off and the distribution tends toward a Gaussian.
For $\Sigma n_{ph} > 30$ the $MT$ distribution can be described by a single Gaussian.} 
   \label{fig:MeanTn}
\end{figure}

Figure~\ref{fig:MeanTz} shows the peak values of the mean time $MT$ distribution for different positions along the fibers.
Since the total distance traveled by light emitted in opposite hemispheres is constant (i.e. the length of the fiber)
one can expect that the mean time $MT$ will remain constant along the fiber and independent of the hit position.
Figure~\ref{fig:MeanTz} shows that indeed this is the case within our resolution,
and that the width of the $MT$ distribution does not change appreciably along the fiber, as well.
Therefore, $MT$ is a good observable for timing measurements,
and $MT$ does not depend on the hit position
(i.e. it does not require to take into account the light propagation time in the fiber).

Figure~\ref{fig:MeanTn} shows the dependence of the mean time $MT$ distribution on the number of detected photons $\Sigma n_{ph}$.
As $\Sigma n_{ph}$ increases, as it has also been the case for the time-difference $\Delta T$ distribution,
the tail on the right of the peak in the $MT$ distribution dies off, and the $MT$ distribution tends toward a narrower Gaussian distribution,
especially for the NOL-11 fiber.
For $\Sigma n_{ph} > 30$ the achievable time resolution $\sigma_{MT}$ with the NOL-11 fiber is $\simeq 220~{\rm ps}$,
while it is $\simeq 240~{\rm ps}$ for the SCSF-78 fiber.
To be noted that the peak positions of the $MT$ distributions (Figure~\ref{fig:MeanTz}) do not move with $\Sigma n_{ph}$
(i.e. the amplitude of the signal).
This indicates that our timing algorithm is not affected by the amplitude of the signal;
in other words we do not observe an amplitude driven time walk.
Therefore, there is no need to apply an amplitude correction to the time measurement.

\section{Discussion}
\label{sec:disc}

In summary, we have studied the timing properties of various blue-emitting scintillating fibers.
The best timing performance has been achieved with the NOL-11 fiber,
while the SCSF-78 fiber gives a slightly higher light yield.
The achievable time resolution with a $\sim 500~\mu{\rm m}$ thick NOL-11 fiber setup is around 250~ps.
With the high yield fibers a detection efficiency close to 100\% can be achieved.
The low light yield fibers are not a good choice nor for timing nor for efficiency.
For different detector thicknesses or different number of fiber layers
one can scale up or down the results reported here.

From this study it is evident that, in order to achieve the best timing,
it is not only important to minimize the fiber's scintillation light decay time $\tau$,
but also to maximize the number of detected photons $n_{ph}$.
This can be achieved, for instance, by increasing the thickness of the SciFi detector
by staggering more fiber layers,
which, however, is not always possible, if one is limited by the material budget of the detector. 
Extrapolating from Table~\ref{tab:MPV}, already with 6 (or 8) staggered fiber layers
one can expect to reach a sufficiently large light yield to be very little sensitive on the fiber's scintillation decay time.
Another possibility for small size detectors of $\mathcal{O}{(10~{\rm cm})}$
to achieve a high scintillation light yield would be to increase the concentration of the {\it activator} dyes. 
However, long attenuation lengths can only be achieved by keeping the concentration of
the {\it spectral shifter} relatively low.

Reading out the fibers at both ends is a must for timing applications.
The mean time $MT$ is a good observable for time measurements, since it does not depend on the particle's crossing point.
We found that the scintillation light propagates in the fibers at half the speed of light in vacuum,
i.e. $v_{\rm fiber} \sim 0.5 \times c$.
We also did not observe edge effects when approaching the fiber's end
as one might have expected because of a different geometrical aperture close to the fiber's end.  

The asymmetric shape of the mean time $MT$ distribution can be mitigated in a likelihood analysis of the time information,
which takes into account the asymmetric shape of the $MT$ distribution,
like in a time of flight based particle identification (PID) analysis, or in coincidence measurements.
The analysis could be further improved by using also the signal amplitude information,
which however is not always available, when using ASICs design to record only the time information.

The results on the timing properties of scintillating fibers presented here complement existing data on the characteristics of scintillating fibers
and fill the gap on the performance of scintillating fibers.
We hope that these results will be useful for the development of new SciFi detectors.

\section*{Acknowledgments}
We would like to thank our \mude \xspace -- SciFi colleagues, in particular L. Gerritzen, for interesting discussions.
We thank C. Joram for bringing to our attention the NOL-11 fibers and providing some samples.
We also thank the technical staff in our Department for help with electronics and in preparing the fiber setups.
The authors acknowledge the funding support from the Swiss National Science
Foundation grants no. 200021\_182031.
We thank the Paul Scherrer Institute for providing the test beam facility used for these measurements.


\end{document}